\RequirePackage{fix-cm}
\documentclass[smallextended]{svjour3}       
\smartqed  
\usepackage{graphicx, color}
\begin{document}

\title{Thermopower of a 2D electron gas in suspended AlGaAs/GaAs heterostructures}


\author{M.\ Schmidt 
\and G.\ Schneider 
\and Ch.\ Heyn 
\and A.  Stemmann 
\and W.\ Hansen         
}


\institute{Matthias Schmidt \at
              Institut f\"ur Angewandte Physik und Zentrum f\"ur
Mikrostrukturforschung,\\Jungiusstra\ss e 11, D-20355 Hamburg\\
              Tel.: +49-40-42838-6124\\
              Fax: +49-40-42838-2915\\
              \email{matthias.schmidt@physnet.uni-hamburg.de}           
}

\date{Received: date / Accepted: date}

\maketitle

\begin{abstract}
We present thermopower measurements on a high electron mobility two-dimensional electron gas (2DEG) in a thin suspended membrane. We show that the small dimension of the membrane substantially reduces the thermal conductivity compared to bulk material so that it is possible to establish a strong thermal gradient along the 2DEG even at a distance of few micrometers. We find that the zero-field thermopower is significantly affected by the micro patterning. In contrast to 2DEGs incorporated in a bulk material, the diffusion contribution to the thermopower stays dominant up to a temperature of 7 K until the phonon-drag becomes strong and governs the run of the thermopower. We also find that the coupling between electrons and phonons in the phonon-drag regime is due to screened deformation potentials, in contrast to piezoelectric coupling found with bulk phonons.
\end{abstract}

\section{Introduction}
\label{intro}
While electrical transport properties of a two-dimensional electron gas (2DEG) embedded in thin membranes has been studied before \cite{Blick}, thermal transport and thermopower investigations of such suspended structures are missing so far. However, the determination of the thermopower of low-dimensional electronic systems provides a powerful tool to obtain complementary information to that obtained from a charge carrier transport measurement. As the latter is only capable to determine the scattering time from the electron mobility, the diffusion thermopower depends on the energy derivative of the scattering time $d\tau/dE$ and is thus capable to classify the electron scattering mechanisms being dominant in the electronic system \cite{handbook} \cite{Sankeshwar}. Further, the phonon induced thermopower (phonon-drag) that arises with a thermal gradient in the crystal lattice, allows to classify the kind of electron-phonon coupling. In macro scaled devices, in which indirect lattice heating is used to generate a thermal gradient along the electronic system, the diffusion thermopower is only hard to extrude due to the dominance of the phonon-drag contribution even in the sub 1 K range \cite{handbook} \cite{Ying} \cite{Fletcher1995}. Most of the efforts concentrated on the reduction of the phonon mean free path. Ying et al. \cite{Ying} used a GaAs substrate that was thinned to 100 $\mu$m. They were able to measure the pure thermo-diffuion of a 2D hole gas at temperatures below 100 mK. Fletcher et al. \cite{Fletcher1995} used a heavily doped substrate and measured the phonon drag up to 0.5 K.  So far, the diffusion thermopower was best accessible by means of a direct electron heating technique in small samples, i.e., with contact separation smaller than the relaxation length between the electronic system and the lattice \cite{Buhmann}.\\
In the following, we present thermopower measurements on a micro sized suspended 2DEG structure in which the 2DEG is included in a Hall bar with dimensions in the order of the phonon mean free path.
The samples were grown via solid source molecular beam epitaxy on a (001) substrate and contain a modulation doped heterostructure.
The design we use combines the advantageous properties of drastically reduced heat conductivity of free-standing, thin structures \cite{Fon} \cite{Tighe.apl}  and the high electric conductivity of 2DEGs embedded in a micro scaled Hall bar device. We will demonstrate that the thermal transport of the lattice in the membrane we use is strongly reduced due to the small dimension. This allows us to establish strong thermal gradients along the 2DEG even at distance of several micrometers which would not be present on structures attached to bulk GaAs  \cite{Fon} \cite{Holland}. We observe that the small dimensions strongly affects the thermopower, suppressing the phonon-drag in the 2DEG up to temperatures of 7 K, which makes the diffusion thermopower clearly measurable. 
\begin{figure}
\centering
\includegraphics[width=0.5 \textwidth]{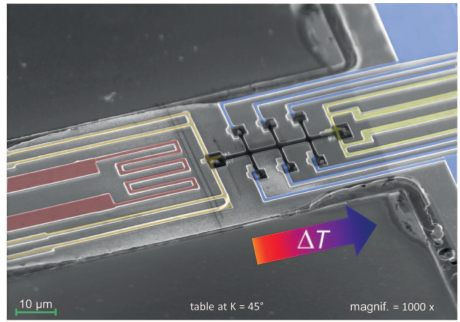}
\caption {Scanning electron micrograph of a suspended, 320 nm thick, 40 $\mu$m wide and 100 $\mu$m long GaAs membrane with a Hall bar device composed of a GaAs/AlGaAs heterostructure containing a 2DEG. The inset shows a schematic drawing of the membrane with color coded features. Micro heater (red) and thermometers (green) enables the application and measurement of the temperature drop directed along the Hall-bar (gray).}
\label{fig:SEM-image}
\end{figure}

\section{Experiment}
The sample design aims at a drastically reduced heat transport along a micro scaled structure including a two-dimensional electron gas so that it becomes possible to establish a strong thermal gradient along the electronic structure of only a few dozens of micrometer in size and to measure the associated thermopower. We use molecular beam epitaxy to grow a modulation-doped AlGaAs/GaAs  high electron mobility transistor (HEMT) structure. 
By the use of e-beam lithography, we defined a 34 $\mu$m long Hall bar, micro-sized thermometers and a micro heater in several lithographic steps as shown in Fig. 1. Segregated AuGeNi supplies Ohmic contacts to the 2DEG and also thermally anchors the electronic system to the lattice temperature. For metallization of the heater and the thermometers we have chosen to evaporate 40 nm AuGe with a small content of Ni (approx. 5 \%). The extra Ni in the metallization is intended to cause a Kondo effect in the leads which results in an increase in the resistivity for low temperatures so, that even temperatures below 5 K can be clearly detected with the thermometers. The thermometers are designed in four-point geometry and measure the temperature at the opposite ends of the Hall bar with the 2DEG which we defined by a mesa etching step. An electrical current through the micro heater feeds Joule heat into the crystal lattice. We suppose it will create a one-dimensional temperature gradient along the membrane, which is thermally anchored at its suspension points. To exclude parasitic influences of the chosen metal on the measurement of the thermopower or transport properties, we fabricated a similar sample with pure Au leads and repeated some of the measurements. The results do not show significant differences.

\section{Thermal conductivity} \label{thermal-cond}
\begin{figure}
\centering
\includegraphics[width=0.5 \textwidth]{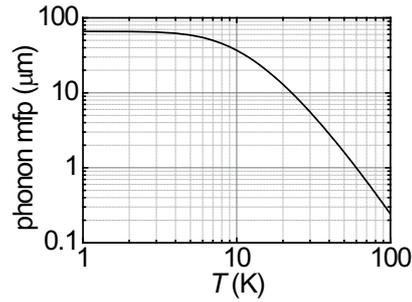}
\caption{Temperature dependence of an estimated phonon mean free path in the membrane indicates that $\kappa$ is essentially limited by the dimensions of the membrane. }
\label{fig:mfp}
\end{figure}

The measurements were performed in a cryostat with variable temperature inset. We used a slow sweep of the ambient temperature to calibrate the resistance thermometers at thermal equilibrium conditions before the measurement. The calibration curves (not shown here) exhibit a minimum of the electrical resistance at a temperature around 80 K, that we attribute to magnetic order effects due to the small extra amount of Ni that we added in the metallization of the AuGe thermometer and heater material. This way, we were able to clearly determine the temperature in the range, where the temperature coefficient of a pure, non magnetic, metal would be too small to obtain accurate values. To measure the thermal conductivity, we apply a constant amount of Joule heating to the heater in the central region of the membrane and measure the temperature drop with the two thermometers over a distance of 34 $\mu$m.  We estimate the phonon mean free path as
\begin{equation}
	\Lambda =\frac{3\kappa}{4cC_{ph}},
	\label{mfp}
\end{equation}
with the measured thermal conductivity $\kappa$, the average phonon group velocity $c$ and a calculated Debye phonon heat capacity $C_{ph}$. Figure \ref{fig:mfp} shows the evolution of the calculated phonon mfp in a temperature range from  1  K to 100 K. The phonon mfp is determined by the membrane geometry and thus temperature independent below $T= 10$ K, indicating the dominance of phonon-boundary scattering. Also due to the small dimension of the membrane, the reduction of the mfp for $T> 10$ K is not attributed to phonon-Umklapp scattering as it would be expected in bulk material \cite{Holland}, but due to the scattering with defects and impurities. A similar observation has been discussed in \cite{Fon} and \cite{Tighe.apl} in detail. 

\section{Thermopower}
The low thermal conductivity of the membrane is used to establish strong temperature gradients over just a few dozens of micrometer distance. With this thermal gradient, we determine the thermopower $S$ by measuring the thermovoltage of a 34 $\mu$m long two dimensional electron gas in a HEMT structure that is located between the thermometers and has an Ohmic contact to each of the thermometers (see Fig. \ref{fig:SEM-image}).

\subsection{Zero-field thermopower}\label{ZeroThermopower}
\begin{figure}[tb]
\centering
\includegraphics[width=0.6 \textwidth]{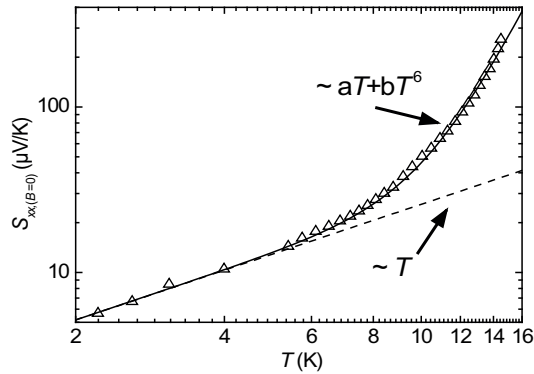}
\caption{Temperature dependence of the zero field thermopower. The dashed line shows the diffusion thermopower calculated via the Mott formula, the solid line corresponds to the total thermopower with the phonon-drag contribution $S_{ph}$ calculated as $S=S_d+S_{ph}=aT+bT^6$.}
\label{fig:Thermopower}
\end{figure}

 The measurements were performed in a temperature range from 2.7 K to 14~K in absence of a magnetic field. The measured zero-field thermopower as shown in Fig. \ref{fig:Thermopower} arises from a combination of two independent contributions. One is the diffusion thermopower $S_d$, resulting from higher energy charge carriers diffusing from the hot to the cold end side. The other is the phonon-drag, arising from a momentum transfer from phonons to electrons when the phonon wind transports heat through the lattice. The total thermopower can be expressed as $S =S_d+S_{ph}$ \cite{handbook}. The diffusion part of the thermopower $S_d$ is linear at low temperatures and can be calculated by the Mott formula 
\begin{equation}
	S_{d}=- \frac{\pi^2k^2_B}{3e}\frac{T}{E_F}(p-1) \label{zero-thermopower}, \label{motteqn}
\end{equation}
which depends on the ratio of the temperature to the Fermi temperature $T_F=E_F/k_B$ measured relative to the bottom of the sub-band and the parameter $p$ that takes the energy dependence of the electronic scattering time into account \cite{Sankeshwar} \cite{Karavolas}. 
The dashed line in Fig. \ref{fig:Thermopower} shows the diffusion component of the thermopower calculated with the experimentally determined charge carrier density of $1.3 \cdot 10^{11}$ cm $^{-2}$, corresponding to a Fermi energy of 4.7 meV. The slope gives best agreement, if we assume a power factor of $p=-0.5$.  As calculated by Karavolas and Butcher for a 2DEG on bulk substrate \cite{Karavolas}, the parameter $p$ has a strong dependence on the charge carrier density. According to their calculation, the parameter $p$ takes values between +1.2 to -1.2 in the carrier density range between 5 and $12\cdot 10^{11}$ cm$^{-2}$. It is thus interesting to perform similar calculations for the case of a 2DEG, with a low charge carrier density as we have, incorporated in a thin membrane.
The phonon-drag component of the thermopower is taken to be of the form $S_{ph}\propto \Lambda T^n$, with the phonon mean free path $\Lambda$ and an exponent n being $n=4$ for piezoelectric coupling and $n=6$ for coupling via a screened deformation potential between the phonons and the electrons in the 2D channel \cite{Sankeshwar} \cite{Kubakaddi}. The total thermopower as depicted in the solid line in Fig. \ref{fig:Thermopower} is calculated with the diffusion thermopower $S_d$ and a phonon-drag part with a constant phonon mean free path and an exponent of $n=6$. Note that the calculation agrees even for higher temperatures, where the mean free path determined from the thermal conductivity becomes temperature dependent as we have shown in Fig. \ref{fig:mfp}. Furthermore, the phonon-drag part in this suspended structure starts to dominate the total thermopower not till temperatures of 7 K which is in contrast to HEMT structures on bulk substrate, where the phonon-drag dominates the thermopower already in the sub-Kelvin range \cite{handbook}. We attribute the late onset of the phonon-drag to the small dimension of our structure being only of the order of the phonon mean free path, so that there are only few electron-phonon scattering events along the sample. 

\subsection{Magnetothermopower}
\begin{figure}[tb]
\centering
\includegraphics[width=0.6 \textwidth]{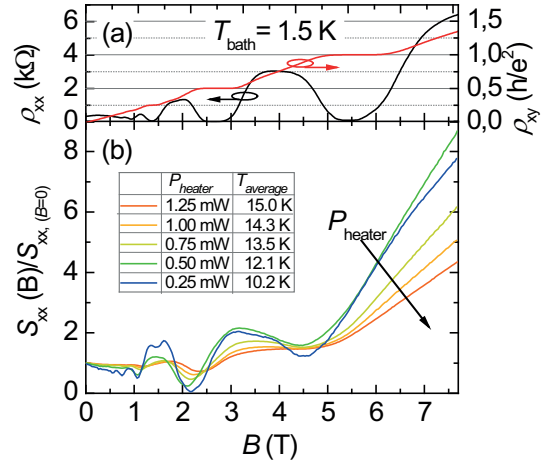}
\caption{(a) Magnetotransport data of the longitudinal (black) and Hall resistivity (red) for comparison with magnetothermopower measurements. (b) Magnetothermopower normalized to the thermopower at $B=0$ in a temperature range from 10.2 K to 15.0 K showing a modulation analogous the the  Shubnikov-de Haas oscillations in the longitudinal resistance. }
\label{fig:MagnetoThermopower}
\end{figure}

The magnetothermopower measurements shown in Fig. \ref{fig:MagnetoThermopower} were performed in a temperature range between 10 K $<T<$ 15 K. In this temperature range phonon drag leads to a rapid increase of the thermopower with temperature. For a better qualitative comparison, the  magnetothermopower $S(B)$ is normalized to the zero-field value measured at each temperature which can be taken from the data shown in Fig. \ref{fig:Thermopower}. The different traces show the characteristic modulation due to an external magnetic field (Fig. \ref{fig:MagnetoThermopower}b). For comparison, magnetotransport data of the longitudinal (red) and the Hall resistance (black) obtained at a bath temperature of 1.5 K with lock-in technique are presented in Fig. \ref{fig:MagnetoThermopower}a. Shubnikov-de Haas oscillations in the longitudinal resistance are well resolved. Filling factor 1 is achieved already at a magnetic field of about 5 T, so that the minimum at $B=5$ T is attributed to the spin-gap. The oscillations in the thermopower correspond to those of the longitudinal resistance as expected.  The Nernst-Ettinghausen signal (not shown here) oscillates with a $\pi/2$ phase shift with respect to $S(B)$ and shows zero crossings at half-filled Landau levels. A slight shift of the minima positions in the thermopower can be associated to a density change during the sample cooling procedure. We note that in the resistance data in Fig. \ref{fig:MagnetoThermopower}a the minimum at $B=5$ T, i.e., of filling factor 1, is weaker in the thermopower. This is explained by the fact that the spin gap is much smaller than the Landau gap.  According to the zero-field data shown in chapter \ref{ZeroThermopower}, we expect the diffusion contribution to the thermopower to be comparable in strength to the phonon-drag at 10 K (see Fig. \ref{fig:Thermopower}), whereas at $T\approx 15$ K the phonon-drag dominates. However, a comparison of the oscillation strengths of the different thermopower traces is complicated due to the relatively high temperatures of the measurements at which quantizing effects are less pronounced.  Both, the diffusion and the phonon-drag thermopower contribute to the oscillations. For a distinction of the diffusion thermopower alone, measurements at lower temperatures are required.

\section{Conclusions}
In conclusion, zero-field thermopower and magnetothermopower in thin suspended membranes are studied. In the experiment 34 $\mu$m long Hall bars are used and the thermal gradient is established by lattice heating.  The thermopower of the 2DEG enclosed in the thin membrane is strongly affected by the small dimensions of the structure. The phonon-drag contribution to the thermopower is suppressed up to 7 K. The $T$-dependence of the thermopower indicates, that in contrast to HEMT structures on bulk substrate, electron-phonon coupling in the phonon-drag regime is dominated by a screened deformation potential. 
\begin{acknowledgements}
The authors thank the Deutsche Forschungsgemeinschaft for financial support via SPP 1386 "Nanostructured Thermoelectrics".
\end{acknowledgements}


\begin{thebibliography}{}
%
%
\bibitem{Blick}
R. H. Blick, F. G. Monzon, W. Wegscheider, M. Bichler, F. Stern, and M. L. Roukes, Phys. Rev. B, \textbf{62}, 17103(2000).

\bibitem{handbook}
For an overview of thermoelectric effects in low dimensional electronic systems, please refer to: B. L. Gallagher and P. N. Butcher, Handbook on Semiconductors, pp.721. Elsevier Science Publishers, Amsterdam (1992) and the review of R. Fletcher, Semicond. Sci. Technol., \textbf{14}, R1 (1999)

\bibitem{Sankeshwar}
N. Sankeshwar, M. Kamatagi, and B. Mulimani, Phys. Status Solidi B, \textbf{242}, 2892 (2005).

\bibitem{Ying}
X. Ying, V. Bayot, M.B. Santos, and M. Shayegan, Phys. Rev. Lett., \textbf{50}, 4969 (1994).

\bibitem{Fletcher1995}
R. Fletcher, P. T. Coleridge, and Y. Feng, Phys. Rev. B, \textbf{52}, 2823 (1995).

\bibitem{Buhmann}
S. Maximov, M. Gbordzoe, H. Buhmann, and L. Molenkamp, Phys. Rev. B \textbf{70}, 121308(R) (2004).

\bibitem{Fon}
W. Fon, K. C. Schwab, J. M.Worlock, and M. L. Roukes, Phys. Rev. B, \textbf{66}, 045302 (2002).

\bibitem{Tighe.apl}
T. S. Tighe, J. M. Worlock, and M. L. Roukes, Appl. Phys. Lett., \textbf{70}, 2687 (1997)

\bibitem{Holland}
M. G. Holland, Phys. Rev., \textbf{132}, 2461 (1964).

\bibitem{Karavolas}
V. C. Karavolas and P. N. Butcher, J. Phys.: Condens. Matter, \textbf{3}, 2597 (1991).

\bibitem{Kubakaddi}
S. Kubakaddi, Phys. Rev. B, \textbf{69}, 035317 (2004).




\end{thebibliography}


\end{document}